\documentclass[10pt,twocolumn,superscriptaddress,showpacs]{revtex4-1}

\usepackage{amsmath}    % need for subequations
\usepackage{graphicx}   % need for figures
\usepackage{verbatim}   % useful for program listings
\usepackage{color}      % use if color is used in text
\usepackage{subfigure}  % use for side-by-side figures
\usepackage{hyperref}   % use for hypertext links, including those to external documents and URLs
\usepackage{units}

\begin{document}

\title{Continuous Loading of a Conservative Trap from an Atomic Beam}
\author{Markus Falkenau}
\affiliation{5. Physikalisches Institut, Universit\"at Stuttgart, Pfaffenwaldring 57 D-70550 Stuttgart, Germany}
\author{Valentin V. Volchkov} 
\affiliation{5. Physikalisches Institut, Universit\"at Stuttgart, Pfaffenwaldring 57 D-70550 Stuttgart, Germany}
\author{Jahn R\"uhrig} 
\affiliation{5. Physikalisches Institut, Universit\"at Stuttgart, Pfaffenwaldring 57 D-70550 Stuttgart, Germany}
\author{Axel Griesmaier}
\affiliation{5. Physikalisches Institut, Universit\"at Stuttgart, Pfaffenwaldring 57 D-70550 Stuttgart, Germany}
\affiliation{Niels Bohr Institute, University of Copenhagen, Blegdamsvej 17, DK-2100 Copenhagen, Denmark}
\author{Tilman Pfau}
\affiliation{5. Physikalisches Institut, Universit\"at Stuttgart, Pfaffenwaldring 57 D-70550 Stuttgart, Germany}

\date{\today}

\pacs{37.10.De, 37.10.Gh, 37.10.Vz, 37.10.Mn, 37.20+j, 67.85.Hj}

\begin{abstract}
We demonstrate the fast accumulation of $^{52}$Cr atoms in a conservative potential from a magnetically guided atomic beam. Without laser cooling on a cycling transition, a single dissipative step realized by optical pumping allows to load atoms at a  rate of $2 \cdot 10^{7}$s$^{-1}$ in the trap. Within less than \unit[100]{ms} we reach the collisionally dense regime, from which we directly produce a Bose-Einstein condensate with subsequent evaporative cooling. This constitutes a new approach to degeneracy where, provided a slow beam of particles can be produced by some means, Bose-Einstein condensation can be reached for species without a cycling transition. 
\end{abstract}

\maketitle

%\section{Introduction}

The development of laser cooling \cite{metcalf02} opened the door to new temperature regimes and was a prerequisite for the advances in the field of atom optics and quantum gases. However, the standard laser cooling mechanism is based on the existence of a closed cycling transition. This restricts the application of laser cooling to merely a few atomic species. 
Whereas for atoms with a rich level structure and molecules, laser cooling is in general technically demanding or impossible; for those, the generation of slow beams of atoms and molecules has become an important experimental technique \cite{meerakker08}. Such beams can be produced for species with an electric or magnetic \cite{mcclelland06, mingwu10, alex07} dipole moment by using pulsed electric or magnetic fields \cite{Lahaye04.1, vanhaecke07, narevicius08, bethlem99}. In these cases, a mechanism that allows to remove the directed kinetic energy from the beam and to transfer the particles into a conservative trap potential - without the need of laser cooling - is highly desirable. \par

A diode that lets atoms pass from a reservoir into a trap volume, but not vice versa, can provide such a mechanism \cite{ruschhaupt04,ruschhaupt07} and has been compared with the realization of Maxwell's pressure demon \cite{thorn08,Price08}. 
The operation of an atom diode can give rise to a pressure gradient, while ideally no heating occurs. The atom number density within the trap volume thus can be dramatically increased with respect to the density of the beam. Such a mechanism paves the way for an alternative to increase the phase space density of the now trapped sample. Neglecting losses, in principle, this scheme could already lead to the formation of a BEC \cite{Klaers10,erhard04,santos01}. Usually, however, inelastic scattering limits the achievable density of such an approach. In this case, given the elastic collisional rate is high enough, the temperature can be decreased further with subsequent evaporative \cite{ketterle96} or demagnetization cooling \cite{fattori06}. \par

For the realization of a one-way barrier, scattering of a single photon, changing the state from an untrapped to a trapped one, is sufficient \cite{schneble03} . This circumvents the need for a closed transition and thus makes it applicable to a larger variety of atomic and molecular species \cite{narevicius09}. \par

\begin{figure}
\includegraphics[scale=0.92]{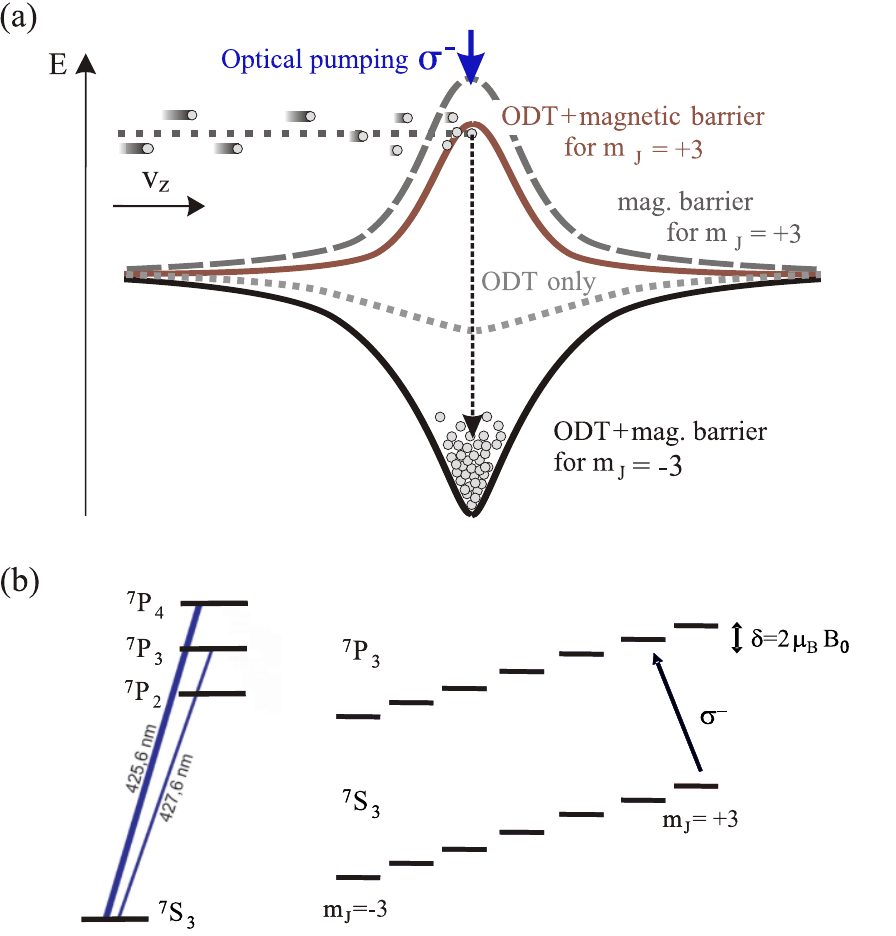}
\caption{\label{fig:potential}\textbf{(a)} Illustration of the dissipative loading scheme from a guided atom beam. A superposition of an ODT and a magnetic field confines the atoms radially (xy-plane). Along the z-direction the hybrid potential can be either repulsive or attractive depending on the atom's magnetic sublevel. Arriving atoms are in the low-field seeking $m_{J}=3$ state. The magnetic field thus acts as a barrier. At the position of the barrier's maximum, the atoms are pumped to the high-field seeking absolute ground state $m_{J}=-3$ while the directed kinetic energy is dissipated. \textbf{(b)} Magnetic substructure of the $^{7}S_{3} \rightarrow \! ^{7}P_{3}$ transition in $^{52}$Cr used for optical pumping.}
\end{figure}

%\section{The Experiment}

In this article, we demonstrate the implementation of a continuous loading scheme based on state dependent potentials and optical pumping only \cite{anoush09}. A hybrid magnetic and optical dipole trapping potential is placed within an atomic beam of $^{52}$Cr atoms. The simplified 1D sketch of this hybrid potential is depicted in Fig. \ref{fig:potential}(a). Atoms are injected longitudinally (along z-axis) into an optical dipole trap (ODT), where they are radially confined. A magnetic potential barrier at the position of the ODT's focus slows the atoms and converts their directed kinetic energy into potential energy. Close to the classical turning point, optical pumping transfers the atoms to a trapped state, while the potential energy is dissipated by spontaneous emission. The trapped state ($^{7}S_{3}$, m$_{j}=-3$) is a dark state for the $\sigma^{-}$ pumping light. 

\begin{figure}
\includegraphics[scale=0.35]{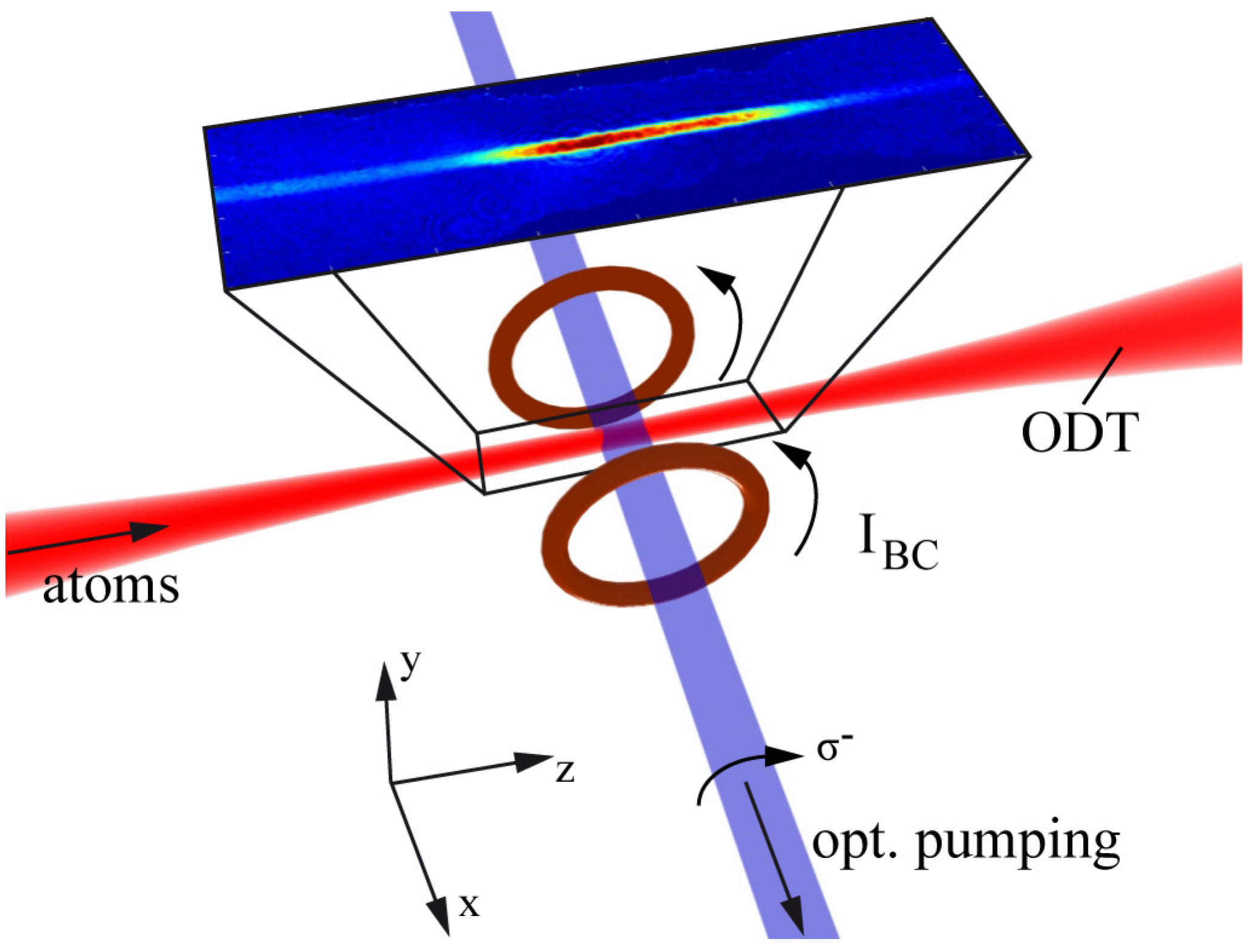}
\caption{\label{fig:schema}Illustration of the experimental realization. The atoms travel along the z-axis towards the focus of the ODT. The atomic beam is radially confined by the ODT laser beam. The magnetic field barrier is created by a pair of coils, fed by a current $I_{\text{BC}}$. An optical pumping beam is passing through the center of the coils and pumps the atoms into the lowest Zeeman sublevel. Absorption images are taken by illuminating the cloud along the y-axis.}
\end{figure}

\begin{figure}
\includegraphics[scale=0.4]{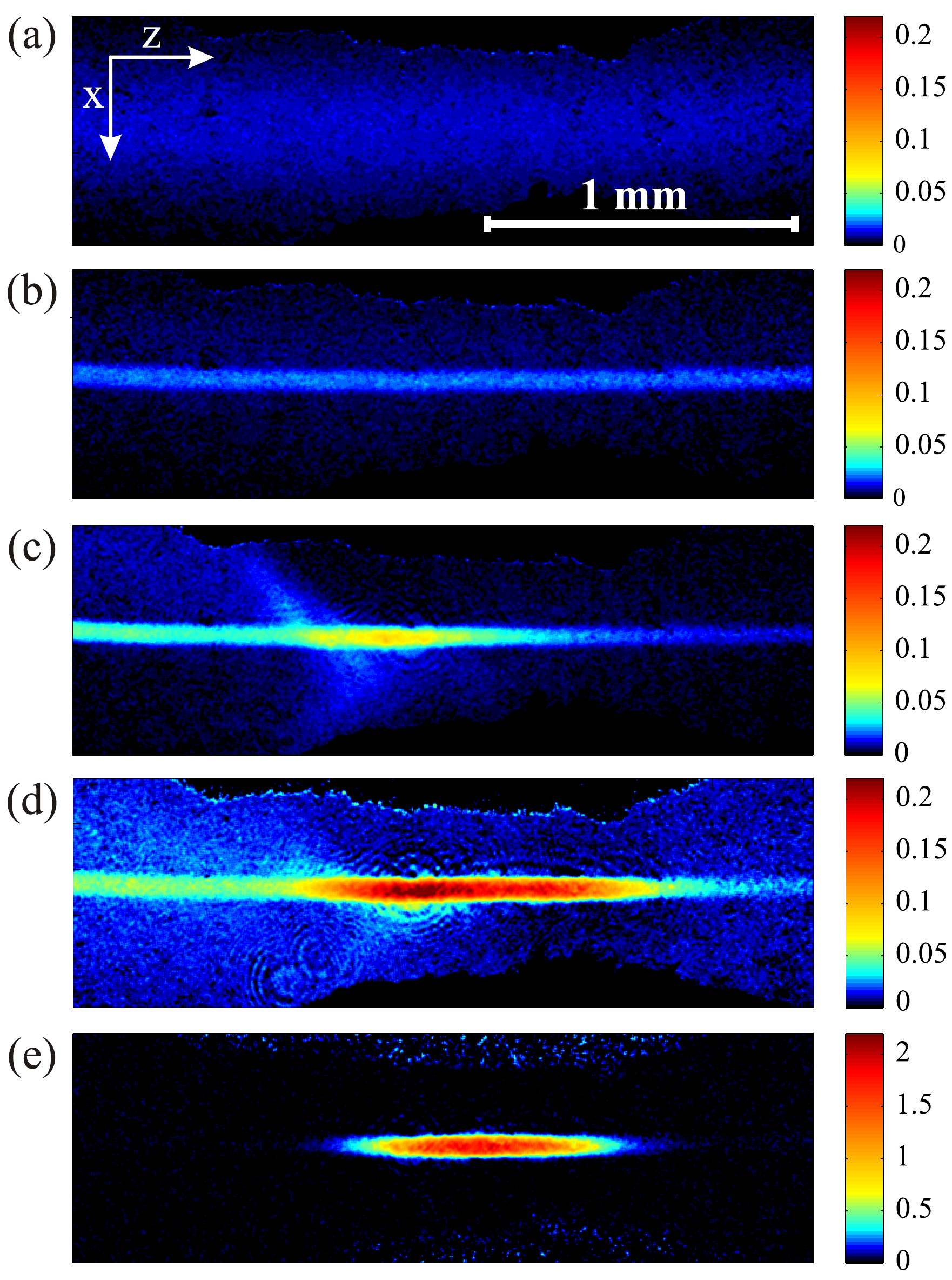}
\caption{\label{fig:beam} In trap absorption images at the position of the cloud.  \textbf{(a) Magnetic guide.} A picture of the magnetically guided atomic beam.  \textbf{(b) Magnetic guide + ODT.} Superimposing the ODT laser beam on axis with the magnetically guided atoms a large fraction of the atomic beam is funneled into the ODT potential.  \textbf{(c) Magnetic guide + ODT + magnetic barrier.} Adding the magnetic barrier the atoms are slowed. Part of the atoms has enough energy to cross the barrier, the rest is reflected.  \textbf{(d) Magnetic guide + ODT + magnetic barrier + optical pumping.} Adding the optical pumping beam completes the dissipative loading mechanism. Atoms accumulate in the trap.  \textbf{(e) Fully loaded trap } After a loading time of \unit[100]{ms} the atomic beam is magnetically shifted to the side, ending the loading process. }
\end{figure}

We carry out the experiment by overlapping an ODT on axis with a continuously loaded magnetic guide. The ODT is created by a laser beam at \unit[1070]{nm} wavelength and \unit[80]{W} power focussed to a waist of \unit[30]{$\mu$m}. The magnetic guide \cite{alex07,axel09} consists of a quadrupole field created by four rectangularly arranged Ioffe bars with a spacing of \unit[46]{mm} and a length of about one meter. The bars carry a current of \unit[180]{A} in alternating directions. The resulting  magnetic quadrupole field has a gradient of \unit[14]{G/cm} and provides radial confinement for atoms in the lowfield seeking substates. The magnetic guide is continuously loaded from a moving molasses MOT (MMMOT) that is operated within the guide field. A compressed region in the magnetic guide with additional transverse laser cooling allows for a further increase of the phase space density of the beam \cite{anoush10}.\par

\begin{figure*}
\includegraphics[scale=0.7]{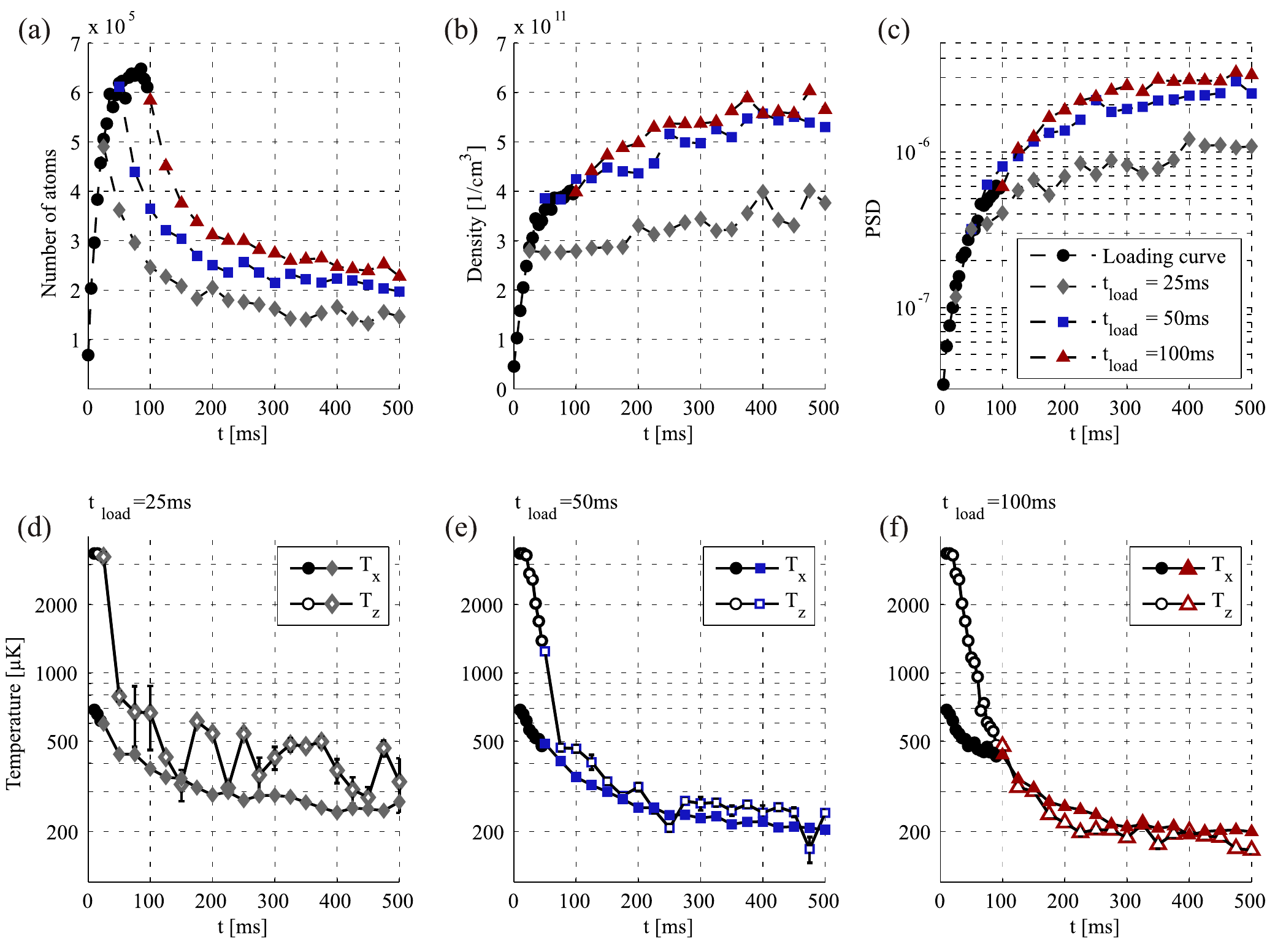}
\caption{\label{fig:daten} \textbf{Evidence for collisions in the loaded ODT}. The graphs show data obtained during trap loading and holding from in trap and time of flight absorption pictures. After loading times $t_{\text{load}}$ of \unit[25]{ms}, \unit[50]{ms} and \unit[100]{ms} respectively, the loading process is stopped. (a) Number of trapped atoms as a function of loading and holding time. (b) Density as a function of time. (c) Phase space density as a function of time. (d)-(f) Evolution of the radial and longitudinal temperature $T_{x}$ and $T_{z}$ respectively for different loading times $t_{\text{load}}$. The shaded areas emphasizes the loading time.}
\end{figure*}

Overall, the MMMOT in combination with the transverse laser cooling provides a continuous beam of $^{52}$Cr atoms with a flux of about \unit[$10^{8}$]{atoms/s} at a velocity of $v=$\unit[1]{m/s}. The radial temperature is below \unit[65]{$\mu$K}, the longitudinal temperature in the moving frame is below \unit[400]{$\mu$K}. The atomic beam is depicted in Fig \ref{fig:beam} (a). The atoms propagate along the z-axis. In-situ absorption images shown here are taken perpendicular to the propagation direction (see Fig \ref{fig:schema}).  \par 

At the position of the focus, a pair of coils is centered as drawn in Fig \ref{fig:schema}. Their diameter and spacing is chosen to be \unit[1]{mm} such that the characteristic decay length of the magnetic field is shorter than the Rayleigh length of the ODT. Therefore the ODT can provide sufficient radial confinement for funneling in the vicinity of the barrier field. The current applied to the coils produces an effective barrier potential for the atom that is equal to their kinetic energy. 
	\[\frac{1}{2}m v^{2} = g_{j} m_{j} \mu_{B}\left|\vec{B}_{\text{max}}\right| + U_{ODT}
\]
In our case, a field maximum of about \unit[10]{G} is needed to stop the atoms. This corresponds to a current of $I_{\text{BC}}=$ \unit[1]{A} for the referred coil configuration. An optical pumping beam resonant to the $^{7}S_{3} \rightarrow \! ^{7}P_{3}$ transition at \unit[427.6]{nm} is aligned through the center of the coils. It has a radial diameter of \unit[100]{$\mu$m} The $\sigma^{-}$ polarized beam pumps the atoms to the lowest Zeeman sublevel. The m$_{j}=-3$ state is both a dark state for the pumping light and trapped in the hybrid potential.\par

Properties of the trapped atomic cloud are extracted from in-situ and time of flight absorption images. A \unit[30]{$\mu$s} light pulse resonant to the $^{7}S_{3} \rightarrow \! ^{7}P_{4}$ transition illuminates the atomic sample perpendicular to the ODT beam. The shadow of the cloud is imaged onto a CCD camera as illustrated in Fig \ref{fig:schema}. Our experimental sequence consists of a loading and successive holding of the trap. We determine  the longitudinal and radial temperature, $T_{z}$ and $T_{x}$, respectively, as well as the atom number $N$ and average density $\rho$ of the trapped atoms. From these numbers, the phase space density is deduced. During the holding phase, the loading process is ended both by shuttering the optical pumping beam and magnetically shifting the atomic beam away from the trap region. For each combination of loading and holding, a time of flight series is taken. \par

Fig \ref{fig:daten} (a) shows the number of trapped atoms as a function of loading and holding time $t_{\text{load}}$ and $t_{\text{hold}}$. The black dots correspond to the loading process. The gray, blue and red points show the evolution of the atom number $N$ while holding, after the trap was loaded for \unit[25]{ms}, \unit[50]{ms} and \unit[100]{ms} respectively. A loading rate of \unit[$2 \cdot 10^{7}$]{atoms/$s$} is achieved. Saturation occurs at about $N \approx 6 \cdot 10^{5}$ after a loading time of \unit[50]{ms}. \par

The atom number density over time is plotted in Fig \ref{fig:daten} (b). During the loading phase, a peak density of \unit[$4 \cdot 10^{11}$]{\text{atoms}/cm$^{3}$} is achieved. While holding the trap, the density increases further, up to about \unit[$6 \cdot 10^{11}$]{\text{atoms}/cm$^{3}$}, which is an indication for plain evaporation. This fact is also proven by Figs \ref{fig:daten} (d)-(f) where longitudinal $T_{z}$ and radial $T_{x}$ temperatures are shown for the respective datapoints. The nonuniform hybrid potential with a depth of about  \unit[5]{mK} longitudinally and \unit[1]{mK} radially as well as the directed velocity along the z-axis of the impinging atoms attribute to the initial non-equilibrium state. It can be seen, however, that thermalization begins within less than \unit[50]{ms} after the loading process is started. The computed phase space density is shown in Fig \ref{fig:daten} (c). Albeit the decrease in atom number, the PSD is steadily growing. Our equilibrium atom density in trap corresponds to a differential density increase of about three orders of magnitude and provides a collision rate around $\unit[10^{2}]{Hz}$. \par

Successive evaporative cooling is done by ramping down the radial confinement, e.g. the ODT's power. The current through the coil pair $I_{\text{BC}}$, is unchanged and thus a constant trapping frequency along the longitudinal trap axis is maintained during evaporation. In this way, we produce an almost pure chromium Bose-Einstein condensate with $10^{4}$ atoms in about $\unit[5]{s}$, which is significantly faster than in previously published experiments \cite{griesmaier05, beaufils08}. \par

Regarding the evolution of the phase space density (PSD) during the continuous loading process, one can assign three phases. Initially, funneling into the radially tighter confining optical potential compresses the atomic beam in an adiabatic way. The PSD is conserved during this step (Compare Fig. \ref{fig:beam} (a) and (b)). Whereas the second step, between Fig. \ref{fig:beam} (b) and (d), corresponds to the dissipative loading process. Each atom that is loaded into the confining potential has a certain energy. Thus the atom number density rises, while the temperature of the trapped atoms remains constant. Thirdly, once the collisional regime is reached, evaporation starts lowering the temperature (Fig \ref{fig:daten} (f)), while loading is still in progress. \par

Considering the limitations of the presented scheme, we observe, that the final atom number and density is robust against small variations in the experimental parameters.  Most likely, inelastic excited state collisions, caused by reabsorption \cite{castin98,santos01} limit the performance of the loading process. The trapped atoms are in a dark state for the pumping light but this state is not dark for spontaneously emitted photons of other atoms during the optical pumping. \par

%\section{Conclusion}

In summary, we have demonstrated a continuous loading mechanism for an ODT from an atomic beam on a sample of chromium atoms. The dissipation by a single optical pumping step effectively acts as a one way valve into our trap from a slow continuous beam. We achieve a final equilibrium density and atom number that is high enough to reach degeneracy by successive evaporative cooling. The existence of a closed transition is not a requirement for this scheme. The application of the loading scheme might ease experiments on ultracold molecules and species which cannot be laser cooled to the collisional regime, but can be slowed down to form a continuous beam. \par

We acknoledge funding from the Deutsche Forschungsgemeinschaft (DFG) within the SFB/TRR 21 and thank A. Aghajani-Talesh for his contributions in the early stages of the experiment. T.P. acknowledges fruitful discussions with J. Thywissen.

%\bibliography{bibfile}

%

\end{document}